\documentclass[final,3p,times,twocolumn]{elsarticle}

\usepackage{amssymb}

\usepackage{verbatim}
\usepackage{makecell}

\usepackage{graphicx}
\usepackage{subfigure}
\usepackage{threeparttable}
\graphicspath{{figure/}}

\begin{document}

\begin{frontmatter}



\title{Attack Tactic Identification by Transfer Learning of Language Model}


\author[inst1]{Ling-Hsuan Lin}
\ead{108356038@nccu.edu.tw}
\affiliation[inst1]{organization={Management Information Systems, National Chengchi University},
            addressline={No.64, Sec.2, Zhinan Rd., Wenshan Dist.}, 
            city={Taipei},
            postcode={11605}, 
            country={Taiwan}}

\author[inst1]{Shun-Wen Hsiao}
\ead{hsiaom@nccu.edu.tw}


\begin{abstract}
Cybersecurity has become a primary global concern with the rapid increase in security attacks and data breaches. Artificial intelligence is promising to  help humans analyzing and identifying attacks. However, labeling millions of packets for supervised learning is never easy. This study aims to leverage transfer learning technique that stores the knowledge gained from well-defined attack lifecycle documents and applies it to hundred thousands of unlabeled attacks (packets) for identifying their attack tactics. We anticipate the knowledge of an attack is well-described in the documents, and the cutting edge transformer-based language model can embed the knowledge into a high-dimensional latent space. Then, reusing the information from the language model for the learning of attack tactic carried by packets to improve the learning efficiency.
We propose a system, PELAT, that fine-tunes BERT model with 1,417 articles from MITRE ATT\&CK lifecycle framework to enhance its attack knowledge (including syntax used and semantic meanings embedded). PELAT then transfers its knowledge to perform semi-supervised learning for unlabeled packets to generate their tactic labels. Further, when a new attack packet arrives, the packet payload will be processed by the PELAT language model with a downstream classifier to predict its tactics. In this way, we can effectively reduce the burden of manually labeling big datasets. In a one-week honeypot attack dataset (227 thousand packets per day), PELAT performs 99\% of precision, recall, and F1 on testing dataset. PELAT can infer over 99\% of tactics on two other testing datasets (while nearly 90\% of tactics are identified).

\end{abstract}

%
%



\end{frontmatter}


\section{Introduction}
\label{sec:introduction}



Advanced Persistent Threat (APT) was a term created by United States Air Force (USAF) analysts in 2006 \citep{apt}, a new breed of threat over the past decades which was highly-organized and well-planned attack against a specific target for a prolonged period. Such attackers' motivations are typically economic or political. Researcher of the global cyber economy, Cybersecurity Ventures, estimates damage losses caused by cybercrime in the next five years will grow at the rate of 15\% per year, reaching \$10.5 trillion U.S. dollars annually by 2025, compared with \$3 trillion in 2015 \citep{CybersecurityVentures}. With the rapidly increasing security attacks and data breaches, cybersecurity is an essential issue of concern to the global priority.

Since there are all kinds of attacks globally, experts have developed unified standards to define different stages or categorize attacks, and organizations share threat information in various report exchange formats. With these results, everyone can describe attack behaviors and intentions in common languages, such as MITRE ATT\&CK lifecycle framework, which describes an attack by tactics, techniques, and procedures (TTPs) \citep{mitreweb}.

Many efforts are done on post-mortem analysis. One is to make cyber threat intelligence (CTI) on attack events. CTI is evidence-based knowledge about existing or emerging threats \citep{cti} to learn the attacker's intention and correlation between attacks. 
However, analysis report relies on case-by-case doing. When a new attack is detected, observing which report the attack is similar to is required to figure out the attacker's target. Researchers \citep{ttpdrill,multisource,paragraphbased,rcatt} proposed automated analysis methods on security reports to reduce human participation. We further discuss these methods in Section \ref{sec:related}.
Although existed efforts on automated analyzing security reports could soothe the labor-intensive matter of attack event in-depth analysis, time-consuming remains the point. In particular, the number of reports output is far less than emerging attacks and numerous complex malicious activities. The speed of defending could hardly catch up with the speed of new attack generating. Transfer learning uses knowledge in the source domain to improve and optimize the learning effect of the predictive model in the target domain \citep{transferlearning}. Therefore, we believe artificial intelligence could help humans analyze attacks automatically, especially analyzing attack intentions for CTI generation.

Our goal is to map network traffic to TTPs and automatically identify the intention of the attack through an AI-based language model, to improve the problem of time-consuming and laborious involved in cybersecurity defense. Our data is collected from honeypots built by an Internet Service Provider. Since HTTP protocol is the most-used text-based protocol in honeypots, our research scope only focuses on HTTP.
To achieve the goal, we face some challenges. First, learning the intention of an attack defined by MITRE ATT\&CK is non-trivial. Even more, MITRE only provides limited attack-related articles describing TTPs. Consequently, we adopt a language model to learn the intention of attacks. Second, manually looking into the attack packet and guessing its intention is inaccurate. Accordingly, we design an AI-based automatic mechanism to deal with TTP labeling. How do we design an automatic system that fulfills a text-based model containing cyber threat intelligence (e.g., referencing TTPs) to analyze network-based data comes to our research problem.

We propose a transformer-based language model that learns the relationship between tactics (intentions) and attack packets by analyzing the TTP articles on the MITRE website. The fine-tune language model can ``understand'' how to read a packet and embed its TTP information in the output vector. We use BERT \citep{bert} to implement our transformer-based language model and establish a dataset generation process with the help of Wireshark \citep{wireshark}, a world widely-used network protocol analyzer, and Snort \citep{snort}, a free, open-source network intrusion detection system, respectively on packet parsing and logging. The process can generate labeled training data with an unsupervised learning method for training a language model.

The main contributions are summarized as follows:
\begin{enumerate}[a)]

\item We leverage a clustering mechanism on real-world data to generate signatures and suitable labels, which covers 94\% of data, significantly reducing the manual labeling burden.
\item Our language model (PELAT) can output a vector representation that embeds the contents of a packet and its intention (if any). As the result, packets are well-clustered under different algorithms.
\item PELAT demonstrates that it outputs at least one of the classes with a high probability by no ``unknown'' class packets were predicted.
\end{enumerate}


\section{Related Work}
\label{sec:related}

\subsection{Background - Cyber Attack Lifecycle}

The cyber attack lifecycle presents a sequence of cyber attack events from which attackers successfully penetrated a network and exfiltrated data. Kill Chain is a term originally used for military purposes to define the chain of events to a successful attack. Expanding on the kill chain concept, Lockheed Martin developed the Cyber Kill Chain model with seven phases of cyber attack in 2011 \citep{killchain}. These phases in sequence are reconnaissance, weaponization, delivery, exploitation, installation, command and control (C\&C), and actions on objectives. The model identifies and helps prevent cyber intrusion activities. Adversaries must complete progress through all phases to achieve their goals, and defenders can stop adversaries at any stage to break the chain of attack.

Attack activities have been more sophisticated over the years. As a result, MITRE Corporation created a MITRE ATT\&CK framework built on the Cyber Kill Chain in 2013 \citep{mitre}. ``ATT\&CK'' stands for MITRE Adversarial Tactics, Techniques, and Common Knowledge. The ATT\&CK framework is a knowledge base of adversary tactics and techniques based on real-world attack behavior observations, describing the technologies used in each attack stage from the adversaries' perspective. Two matrices included under the ATT\&CK framework are the enterprise matrix and mobile matrix. The enterprise matrix covers platforms on Windows, macOS, Linux, PRE, Azure AD, Office 365, Google Workspace, SaaS, IaaS, Network, and Containers, while the mobile matrix covers Android and iOS platforms. In this paper, we focus on discussing the enterprise matrix.
Tactic depicts adversaries' objectives and ``why'' of an attack technique. Technique represents ``how'' adversaries achieve their objectives and on what kind of platform they may operate. ATT\&CK contains 185 techniques and 367 sub-techniques within 14 tactics. Each tactic has multiple techniques, and a technique can be categorized into multiple tactics. For example, the Scheduled Task/Job technique is used by adversaries for Execution, Persistence, and Privilege Escalation.

\subsection{Language Model - Transformer}

Surpasses predecessors such as recurrent neural networks (RNN), long short-term memory (LSTM), and gated recurrent units (GRU), which show the main disadvantage on model's runtime increase as the input sequence length increases, the current state-of-the-art approach in natural language processing (NLP) is Transformer. Transformer \citep{attention} is a sequence-to-sequence neural network architecture based on a self-attention mechanism, with essentially a stack of encoder and decoder layers. Each encoder consists of a multi-headed self-attention layer and a fully connected feed-forward neural network to map input and attention information into vector representations. Decoder presents both layers as encoder does, with an additional attention layer that helps focus on informative parts of the input sentence and turns the representation into output text. Another advantage is that the multi-head mechanism allows for more parallelism than RNN’s during the training, proving to boost efficiency and computing speed.

BERT \citep{bert} is a bidirectional encoder-only transformer published by Google in 2018, consisting of two steps of pre-training and fine-tuning. The two pre-training unsupervised task includes Masked LM (MLM) and Next Sentence Prediction (NSP). The pre-training corpora are English Wikipedia and BooksCorpus. Google has released BERT models with different combinations of layer numbers (\textit{L}), hidden size (\textit{H}), and self-attention head numbers (\textit{A}), which the two primary models are BERT$_{Base}$ (\textit{L}=12, \textit{H}=768, \textit{A}=12) and BERT$_{Large}$ (\textit{L}=24, \textit{H}=1024, \textit{A}=16). After the release of the BERT model, there have been many model optimizations of training designs and strategies, such as more data training, parameter quantification, weight pruning, and knowledge distillation \citep{roberta,albert,distilbert,tinybert}. The purpose of these model improvements is to shorten the training time without affecting performance or to make the model achieve better results on downstream tasks. BERT can be utilized in many tasks by fine-tuning the pre-trained model. BERT provides four downstream tasks on fine-tuning: sentence pair classification task, single sentence classification task, question answering task, and single sentence tagging task. We can adapt these tasks or custom ones to fit our own applications. To be more specific in the domain and its application, BioBERT \citep{biobert} and ClinicalBERT \citep{clinicalbert} respectively continue pre-train the language representation model with biomedical texts and clinical notes, as opposed to SCIBERT \citep{scibert} pre-trains the model from scratch on computer science domain and broad biomedical domain corpora.

\subsection{TTPs Mapping}
There have been various CTI sharing formats over the past few years. Based on the sharing, researchers had done some mapping methods of automated analysis on cyber threat intelligence reports.

Husari et al. \citep{ttpdrill} created a tool called TTPDrill. They extracted threat actions from cyber threat reports by natural language processing (NLP) and Information Retrieval (IR) techniques, captured the relationships of threats learned from ATT\&CK and CAPEC \citep{capec}, and mapped each threat action with the highest score to techniques, tactics, and a kill chain phase, where tactics were obtained through technique retrieval. They generated the output into Structured Threat Information eXpression (STIX) \citep{stix} format.

Ayoade et al. \citep{multisource} implemented a bias correction method and SVM classifier on classifying kill chain phases and TTPs. The dataset includes worldwide threat reports with manually labeled TTP and already-labeled TTP documents from ATT\&CK website. They scraped and extracted TF-IDF features using NLP techniques from each document. Since threat reports are provided by security organizations around the world, deviation exists between the distribution of training data and the distribution of test data. They utilized a bias correction technique to address the bias and continued training a classifier to get the final result, where they made use of tactics to retrieve techniques.

Thein et al. \citep{paragraphbased} proposed a neural network classifier for analyzing threat intelligence reports. The analysis was paragraph-based, including event information extraction and five of the seven kill chain phase estimation. The event information extraction extracted core feature words of the diamond model \citep{diamond} from each security report paragraph. The chain phase estimation classified kill chain phases on extracted threat information. They manually labeled kill chain phases on every ATT\&CK technique as training data with word embedding after preprocessing and further tested on four security reports.

Legoy et al. \citep{rcatt} developed rcATT, ``reports classification by adversarial tactics and techniques,'' a multi-label text classification model for retrieving ATT\&CK tactics and techniques from cyber threat reports, and a post-processing method allowed improving prediction over time, then output results in STIX format. The multi-label text classification was done using a TF-IDF weighted bag-of-words representation and a binary linear SVM. The post-processing method was an unsupervised method that used confidence scores resulting from each classification type to add tactics or remove techniques for prediction improvement.

\subsection{Packet AI Embedding}
Various deep learning methods are used for malicious traffic detection. Therefore, we did some research on applying embedding methods to its approach.
Hwang et al. \citep{packetlstm} presented a word-embedding and LSTM model to classify packets into a benign or malicious state without any pre-processing flow.
Yu et al. \citep{deephttp} constructed DeepHTTP through Bi-LSTM and attention mechanism to perform HTTP traffic anomaly detection and pattern mining. DeepHTTP learns the feature of content and structure of traffic automatically, which identifies semantics and structure of traffic with the input of embedding vector.
Han et al. \citep{packetattention} introduced a hierarchical attention model which learns information from bytes and packets level with the use of bidirectional GRU to build the representation. They utilized the bytes embedding method to process the network flow as word2vector \citep{word2vec}. Besides, they introduced Flow-WGAN that generates different types of applications by learning features from original data, can evaluate the performance of an intrusion detection method.
Goodman et al. \citep{packet2vec} designed Packet2Vec that utilized N-grams Word2Vec embedding to create vectorized representation for each packet instead of creating hand-crafted features on feature extraction, which domain expertise was not needed. Packet2Vec well behaved on the speed of processing and classification of detecting malicious network activity.

Using embedding in these past works, most of them classified network traffic as benign or malicious—only a few classified advanced types such as port scanning, DoS attack, or web attack. Our work analyzes packets but focuses on the attack process and semantic judgment of what attackers aim to do.

In summary of this section, we compared the past related research \citep{ttpdrill,multisource,paragraphbased,rcatt} and our work PELAT in Table.~\ref{tab:conparison}. PELAT outperforms them on tactic prediction task with accuracy of 94.7\%.

\begin{table*}[t]
\small
\centering
\begin{threeparttable}
\begin{tabular}{|l|l|l|l|l|l|}
\hline
& \textbf{TTPDrill} & \textbf{Ayoade} & \textbf{Thein} & \textbf{rcATT} & \textbf{PELAT } \\
\hline
Training Src. & CTI & ATT\&CK + Symantec & ATT\&CK & security report & ATT\&CK \& ref \\
\hline
Training Rep. & N/A & 169 + 17,600 & N/A & 1,490 & 525 + 892 \\
\hline
Eval. Rep. & 50 & 488 & N/A & 1130 & 142 \\
\hline
Label & Manual & Manual & Manual & Manual & Semi-supervised \\
\hline
Method & \makecell[l]{SVM\\POS tagging}   & TF-IDF   & \makecell[l]{Word\\Embedding}   & \makecell[l]{Word\\Embedding}   & \makecell[l]{Sentence\\Embedding} \\ 
\hline
Goal & TTP \& Kill Chain & TTP \& Kill Chain & Kill Chain & TTP & TTP \\
\hline
Performance & \makecell[l]{\textit{tactic} \\ Prec: 84\% \\Recall: 82\%} &
\makecell[l]{\textit{tactic} \\Acc: 93.6\% } &
\makecell[l]{\textit{kill chain} \\Acc: 65\% \\F1: 67\%} &
\makecell[l]{\textit{tactic} \\ Prec: 79.3\% \\Recall: 12.2\% \\ F$_{\mathrm{0.5}}$: 37.8\%} &
\makecell[l]{\textit{tactic} \\ Acc: 94.7\% \\ Prec: 81\% \\ ROC: 91.5\%} \\

\hline
\end{tabular}
\begin{tablenotes}
    \footnotesize
    \item[*] Prec: Precision, Acc: Accuracy
\end{tablenotes}
\caption{Comparison of related research and PELAT}
\label{tab:conparison}
\end{threeparttable}
\end{table*}

\section{PELAT}
\label{sec:design}

\subsection{Problem Definition and Design}
To train a MITRE ATT\&CK tactic classifier with attack packet as the input is relatively difficult. In this study, we investigate the power of transfer learning that learns the knowledge of MITRE ATT\&CK tactic from CTI reports that already specify the tactic labels, and then transfer the knowledge from the original document space to a mapped packet space for latter tactic classification. Thus, we can leverage the well-labeled CTI reports for learning and expect a better learning result for the tactic classifier on network packets.

Given a set of CTI reports, $R$, and a set of predefined MITRE ATT\&CK tactic classes, $M$ (and $|M| = 14$ while MITRE has 14 tactics). A tactic, $M_i$, has $R_i$ reports, and in total there has $P_i$ paragraphs in the reports of this tactic. Given a pre-trained language model $LM$ (in our study, BERT$_{Base}$\cite{bert} is used), we fine-tuned the model with a downstream task $D$ (a tactic classifier) using the $P_i$ paragraphs of all $M_i$ tactics. The fine-tuned model is $LM^{+}$. Then, given a set of HTTP packet, $H$. Each packet, $H_j$, is represented as a sequence of bytes and they are fed into the $LM^{+}$. The $LM^{+}$ output vector of a packet is $V_j$.

For the semi-supervised learning process (only part of the packets, $H_s$, has labeled (including $|M|$ labels and one `Non-attack' label), and the rest of packets, $H_u$, has no label, i.e., $H_s \cup H_u = H$), a neural network based tactic multiple label classifier $C$ accepts an embedded training packet in $V_s$ as input and output a vector of size $|M|+1$, which is the labeled tactic or `Non-attack'. Then, the unlabeled $V_u$ are fed into the $C$ for inferring their tactic. For some embedded packets in $V_u$, $C$ may not output high enough probability ($th$) for all tactics; therefore, we add an additional label `Unknown' and this packet is labeled as `Unknown'. In total, we have $|M|+2$ classes.

For the transfer learning process, all $H_u$ and $H_s$ and their corresponding labels are used to perform end-to-end training to output the even-more-fine-tuned language model, $LM^{++}$, and the multiple label classifier, $C^{+}$.

\subsection{System Overview}
\begin{figure*}[tbp]
\centerline{\includegraphics[trim = {0mm 80 10 0}, clip, width=0.75\linewidth]{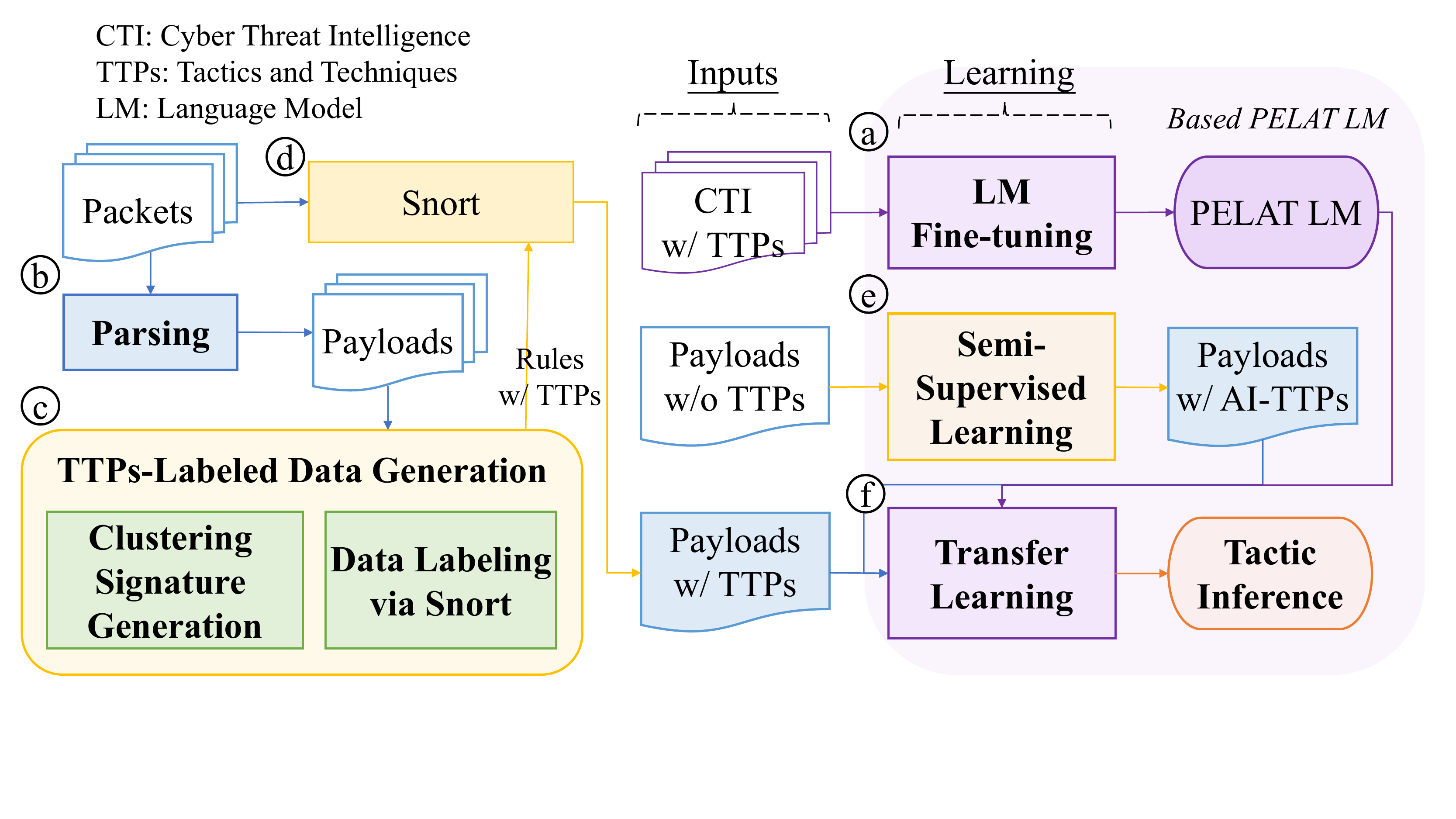}}
\caption{PELAT system overview.}
\label{fig:overview}
\end{figure*}

Figure~\ref{fig:overview} shows the big picture of our designed system PELAT, ``Packet Embedding Method Based Language Model for Attack Lifecycle Knowledge on Tactic Inference,'' with the goal of identifying attack intentions from incoming traffic. First, we parse packets and reserve HTTP payloads, then do some preprocessing for the further process. Next, we designed several modules and a language model that learns cyber threat intelligence. The modules are a dataset-labeling process. The first subprocess contains a clustering signature generation module and a Snort rule module. The second subprocess goes to semi-supervised labeling. We transfer the prior knowledge of ATT\&CK. Hence, we make TTP inference on packets by the PELAT language model.

\subsection{The fine-tuning process of PELAT language model}
\label{sec:languagemodel}
\begin{figure}[tbp]
\centerline{\includegraphics[trim = {10mm 160 60 50}, clip, width=1\linewidth]{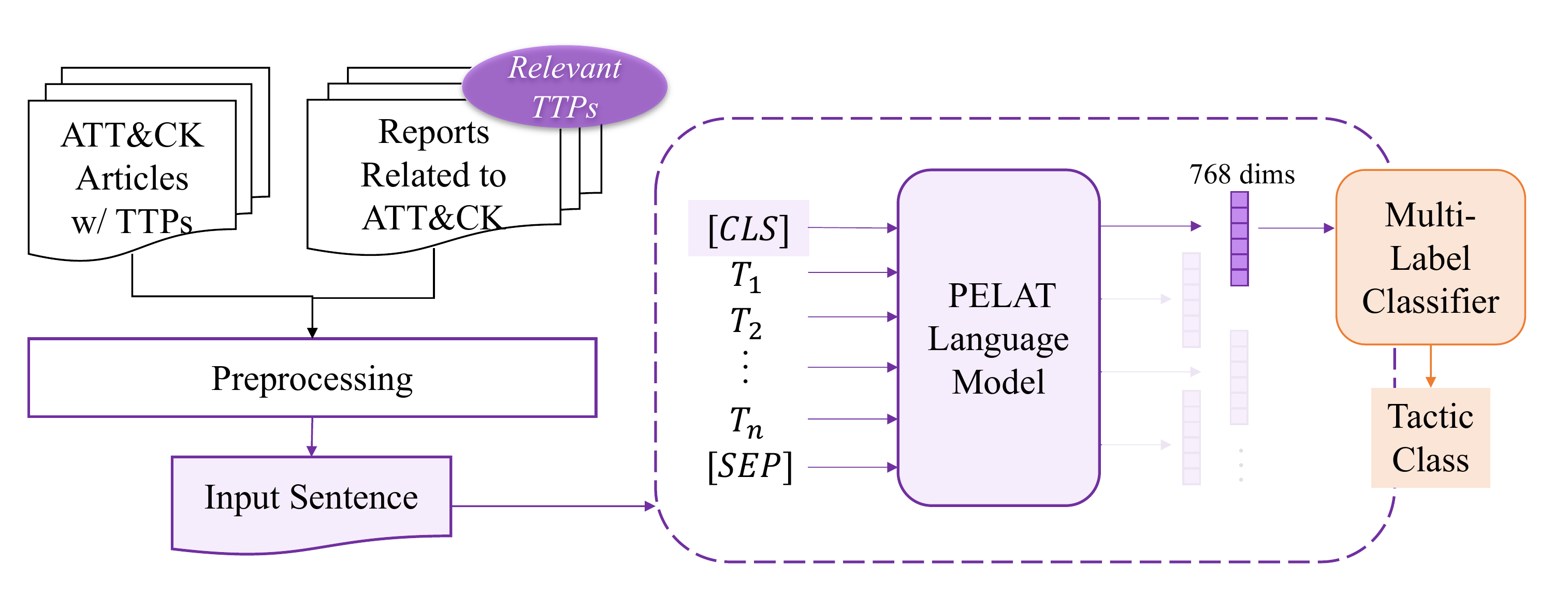}}
\caption{The fine-tuning process of PELAT language model with downstream tactic classifier by labeled CTI reports.}
\label{fig:languagemodel}
\end{figure}

To mark TTP on packets, the model should understand the knowledge of the attack lifecycle (see a in Fig.~\ref{fig:overview}). We select BERT as our language model and select ATT\&CK as the base of attack lifecycle knowledge. We collect articles and reports related to MITRE ATT\&CK, which are limited and lack labeled TTP from websites by web scraping. For preprocessing, we separate long reports into shorter paragraphs to avoid truncating too many words on the restrict of BERT max input length. We then convert the data into a BERT compatible input format and add a classification layer after the transformer encoder layers. Note that an ATT\&CK technique may cover more than one tactic. Due to this ATT\&CK characteristic, we do multi-label classification. The PELAT language model (shown as Fig.~\ref{fig:languagemodel}) can learn cyber threat intelligence and embed TTP information in a multidimensional space.

\subsection{Packet Parsing and Preprocessing}
\label{sec:payloadsparsing}
A network packet is the smallest unit of transmitting data over a network. A packet structure contains information of header and data (aka payload). We parse packets (see b in Fig.~\ref{fig:overview}) and only use the payload (ignore the header) to represent a packet.
In HTTP, a payload content includes HTTP header information, e.g., User-Agent, Accept, Connection, Content-Length, and Host. However, some information that is not important may somewhat be regarded as \textit{noise} in our language model. Accordingly, we remove ``Accept'' and ``Accept-Encoding'' to reduce the burden on the model and make regular expressions for IP Address into ``IP'' and the version of various ``User-Agent'' into ``VERSION''.

\subsection{Labeling for the Semi-supervised Learning}
\label{sec:generation}
\begin{figure}[tbp]
\centerline{\includegraphics[trim = {0mm -30 0 -30}, clip, width=1\linewidth]{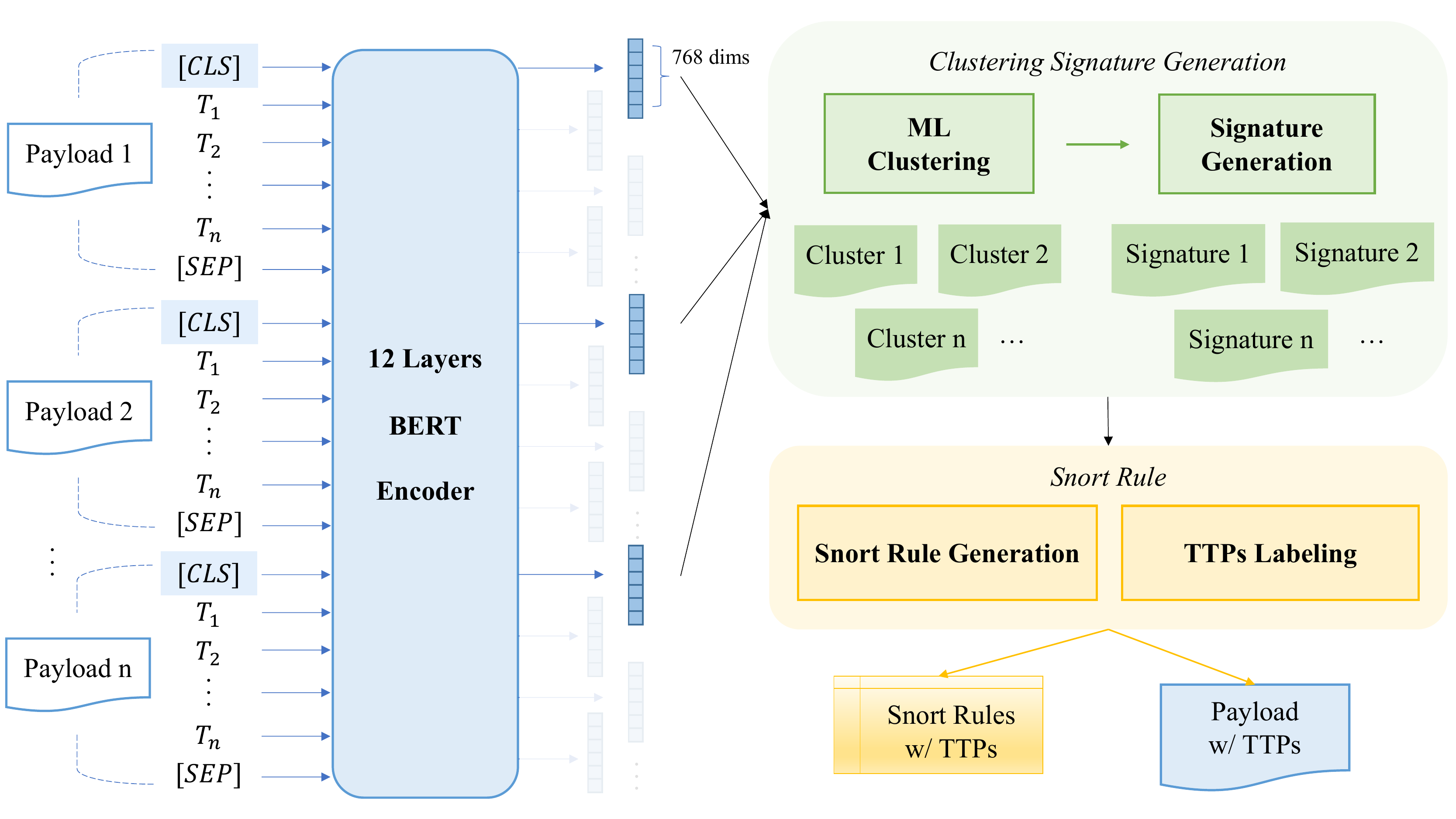}}
\caption{The process of generating TTP-label for semi-supervised learning.}
\label{fig:generation}
\end{figure}

Figure~\ref{fig:generation} shows the process of generating TTP-label for semi-supervised learning. Since the original dataset has no any labels, we adapt a clustering mechanism to cluster similar packet payloads and generate signature and corresponding tactic for each cluster with the help of applying TF-IDF on the clusters. It can efficiently label large portion of the packets by leveraging the cluster. The labeling process (see c in Fig.~\ref{fig:overview}) includes a clustering signature generation module and a Snort rule module. The clustering signature generation module consists of machine learning clustering and signature generation. The Snort rule module consists of generating Snort rules that can match packet payload with signature and output its TTP.

\subsubsection{Clustering Signature Generation}
In this section, our purpose is to extract the important signature in a packet payload. Instead of feature extraction done by machine learning methods or deep learning methods \citep{realtimeextraction,webextraction,cnnextraction}, we take advantage of the language model that keeps semantics and syntactic information to construct vectorized representation on packet payload through 12 layers BERT encoder. This embeds tokens into 768-dimension vectors. For every payload sequence, we reserve the first token [\textit{CLS}] in the last hidden layer of BERT encoder for its representation.

Each payload is scattered in a 768-dimensional latent space after converting payloads into vectors. We considered payloads with similar characters to be high potentially closed to one another. Thus, we apply a machine learning clustering that can minimize the distance within the group and maximize the distance outside the group with the input data of payload representations. For the clustering result, the point closest to the centroid implies the representative of the payload cluster, which also indicates we can generate $n$ signatures from $n$ clusters.

\subsubsection{Data Labeling via Snort}
Now we have numbers of payload signatures. This section describes how to generate TTP labels into packets through signatures. First, we map the signatures into TTP. Second, we look into every payload and determine which signature belongs so that the payload can be labeled to the appropriate TTP. Note that we add an ``Non-attack'' class for the packets that do not belong to attacks. The fastest way to handle millions of payloads is with the help of Snort (see d in Fig.~\ref{fig:overview}). For this, we define Snort rules and execute Snort command that creates log files listing the packet arrival time and its belonged signature. We combine the payloads and logs with the same time value and thus get the initial version of label data.

\subsection{Semi-supervised Learning}
\label{sec:semisup}
Payload signatures generated by the clustering result indicate these signatures are the top $n$ largest or most common signatures in our honeypot. We assume that not all payload signatures are clustered appropriately. A small number of signatures may scatter outside the cluster or fall on the cluster's edge due to their low amount or inconspicuous features. Furthermore, there are signature variants even if the signatures are in the same cluster. We design a semi-supervised labeling process (see e in Fig.~\ref{fig:overview}) to robust our model and more reliable labeled data.
We make inferences on payloads without labels through the attack lifecycle knowledge language model mentioned in Section \ref{sec:languagemodel}. We add an ``Unknown'' class for our assumption. Once our language model infers all the tactic probabilities of a payload less than a threshold, the payload belongs to the unknown class. Otherwise, it belongs to the inferred tactic.

\subsection{Transfer Learning}

In the transfer learning process (see f in Fig.~\ref{fig:overview}), all the packets from the subsection \ref{sec:generation} and \ref{sec:semisup} are labeled. We now can use them to perform end-to-end training to output the final language model and its tactic classifier. The language model can well-represent the text-based payload (with the transferred knowledge from CTI reports) by a vector, while it is then used by the latter neural network to map its corresponding MITRE ATT\&CK tactic. We expect the latent space of the language model (PELAT) can better represent a packet after transfer learning. 

\section{Implementation}
\label{sec:implementation}

\subsection{Dataset}
The honeypots are deployed under five Class C subnets of Taiwan Internet Service Provider Chunghwa Telecom (CHT) to collect real-world packet data. A total of four types of honeypots are deployed: Amun, Cowrie, Dionaea, Glastopf. The deployed honeypots leave logs of the interaction between the outside world and the trapping network, and also retain the original packet data (pcap).

Honeypots indeed interact with the outside world. When receiving external behavior, a honeypot responds within its ability. A pcap file includes inbound and outbound data, where inbound data represents the received packet, and outbound data represents the source IP of a packet from the honeypot address. We assume incoming traffics are malicious activities. We only analyze the inbound data instead to identify what their intention is.

\subsection{Environment}
We have a big data platform consisting of three server clustering and two Network Attached Storage (NAS). The platform runs on Kubernetes with Hadoop Distributed File System (HDFS) storage and Apache Spark. Spark supports distributed computing and is used for large-scale data processing. Our collected honeypot data are stored in NAS. We train our model on one of the servers with Intel i7 CPU, 64GB RAM, 745GB SWAP, and an Nvidia GTX-1080Ti 11GB GPU.

Python 3.7 is our main programming language. We scraped the articles from websites by BeautifulSoup package. For model training, we use CUDA 10.1, Pytorch 1.9.0 PyTorch Lightning framework \citep{pytorchlightning} Transformer library \citep{huggingface}. At last, Scikit-learn \citep{scikitlearn} is used to evaluate the performance.

\section{Evaluation}
\label{sec:evaluation}
\subsection{The fine-tuning process of PELAT language model}
\label{sec:languagemodelCTI}
\subsubsection{Dataset and Preprocessing}
We scraped articles from the MITRE ATT\&CK website and its reference reports. Since not all data we scraped can be referred to as sentences, we did some preprocessing. We parse sentences from articles by Stanza \citep{stanza} and keep sentences that are more than three words. Stanza gives a universal POS tag to every word in a sentence. We set several rules for different POS tag combinations that we believe these combinations are noise.
A total number of 1,417 HTML articles are retained after passing the rules. We assume reference articles are statements related to the technique. We assign labels to the article with the corresponding tactic of the technique. It is possible that an article is referenced by many techniques. To be more tightly, we only reserve articles with less than three tactics. 1,391 remaining articles and the labeled tactic distribution (Table~\ref{tab:dataset} ``CTI Doc'' column) is shown.


\subsubsection{Settings and Result}
We fine-tune BERT$_{Base}$ uncased model with the MITRE ATT\&CK dataset for four epochs and use the Adam optimizer learning rate \texttt{2e-5}, dropout probability of  0.1. Parameter initialization is the same as BERT \citep{bert}. Due to the BERT input length limitation, we divide articles into 18,472 small paragraphs. According to the paragraph-token distribution, the maximum sequence length is 256 tokens with a batch size of 16.

\begin{figure}[tbp]
\centerline{\includegraphics[trim = {0mm 0 0 0}, clip, width=1\linewidth]{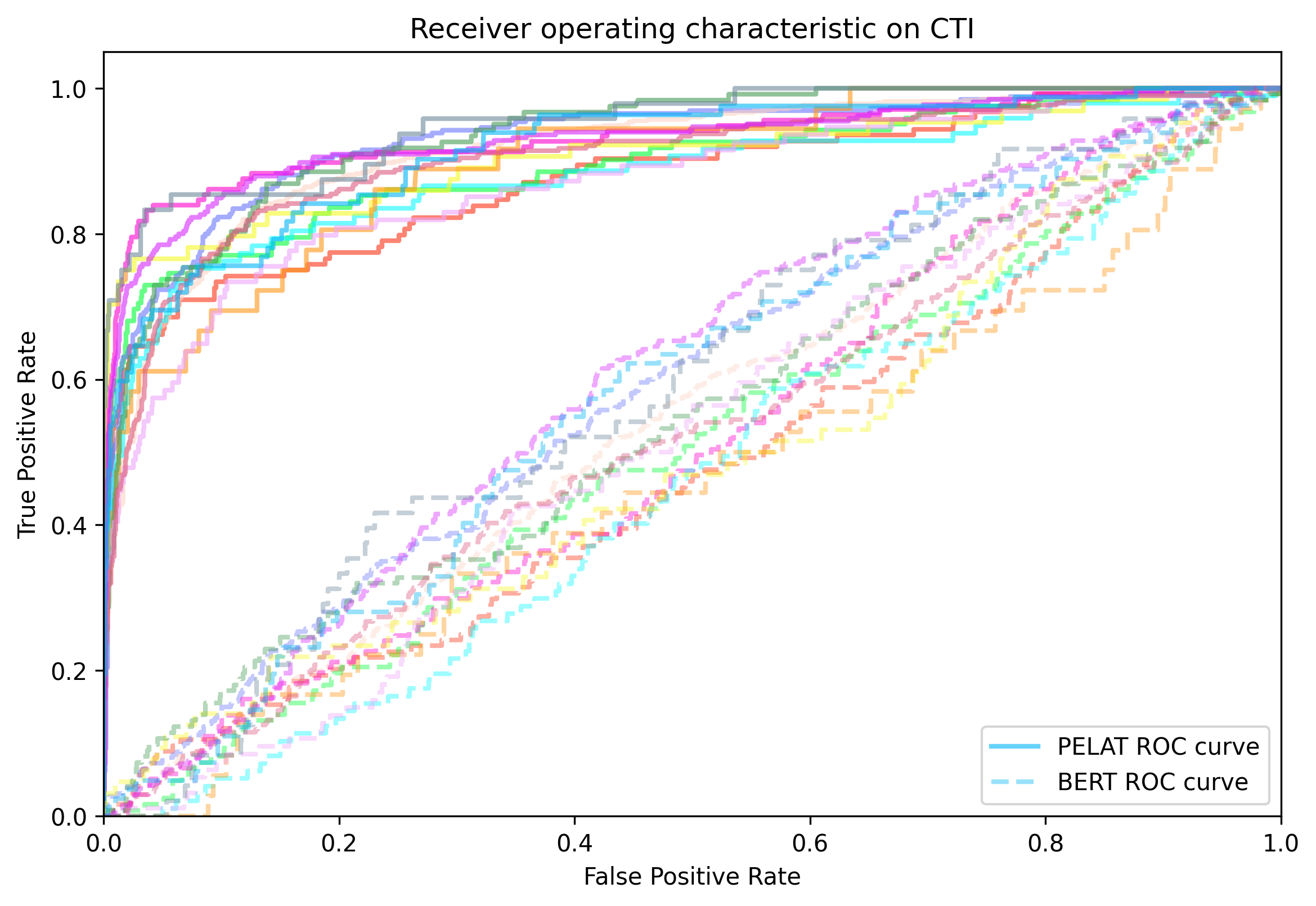}}
\caption{The individual ROC curve of 14-tactic classification using original BERT and trained PELAT embedding.}
\label{fig:roc_article}
\end{figure}

To evaluate the fine-tuned model, we add a single layer forward network with sigmoid activation function to construct a non-linear classifier to predict 14 tactics. The dataset is split into training, test, and validation sets with the ratio of 85\%, 10\%, and 5\%.
Figure~\ref{fig:roc_article} shows the ROC curve of testing dataset on 14 tactics using original BERT and trained PELAT. The solid lines are the ROC curve of PELAT for 14 tactics and the dotted ones are BERT$_{Base}$-ROC.
Table~\ref{tab:article_performance} shows that PELAT has 94.7\% accuracy and 91.5\% ROC, which performs better than BERT$_{Base}$ model.

\begin{table}[tbp]
\small
\caption{The Accuracy and ROC of Language Models for CTI Tactic Classification}
\begin{center}
\begin{tabular}{lcc}
\hline
\textbf{Model}  & \textbf{Accuracy} & \textbf{ROC} \\
\hline
BERT$_{Base}$  & 91.6\% & 74.4\% \\
PELAT       & \textbf{94.7\%} & \textbf{91.5\%} \\
\hline
\end{tabular}
\label{tab:article_performance}
\end{center}
\end{table}

\subsection{Payload Embedding}
\subsubsection{Dataset and Preprocessing}
Honeypots captured a total of 621,365 HTTP inbound packets in two days (the day before and on the day of 2020 Taiwan’s presidential election). We parse payloads from packets and do the preprocessing mentioned in Section \ref{sec:payloadsparsing}. 



\subsubsection{Embedding Result and Cluster Analysis}
\begin{figure}[t]
\centering
\subfigure
[Colored by K-means cluster]
{\includegraphics[width=0.8\linewidth]{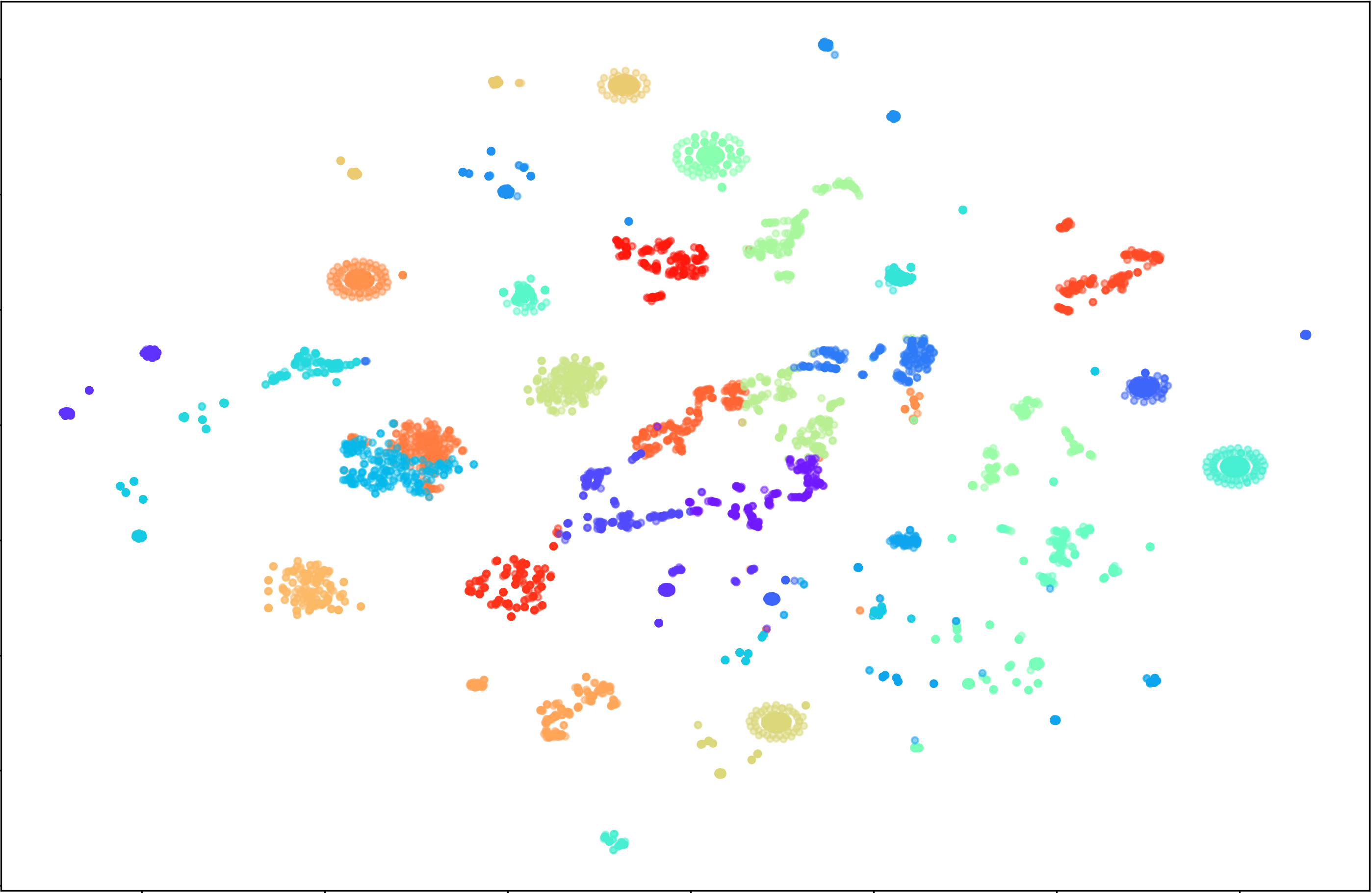}}
\subfigure[Colored by tactic label]{\includegraphics[width=0.8\linewidth]{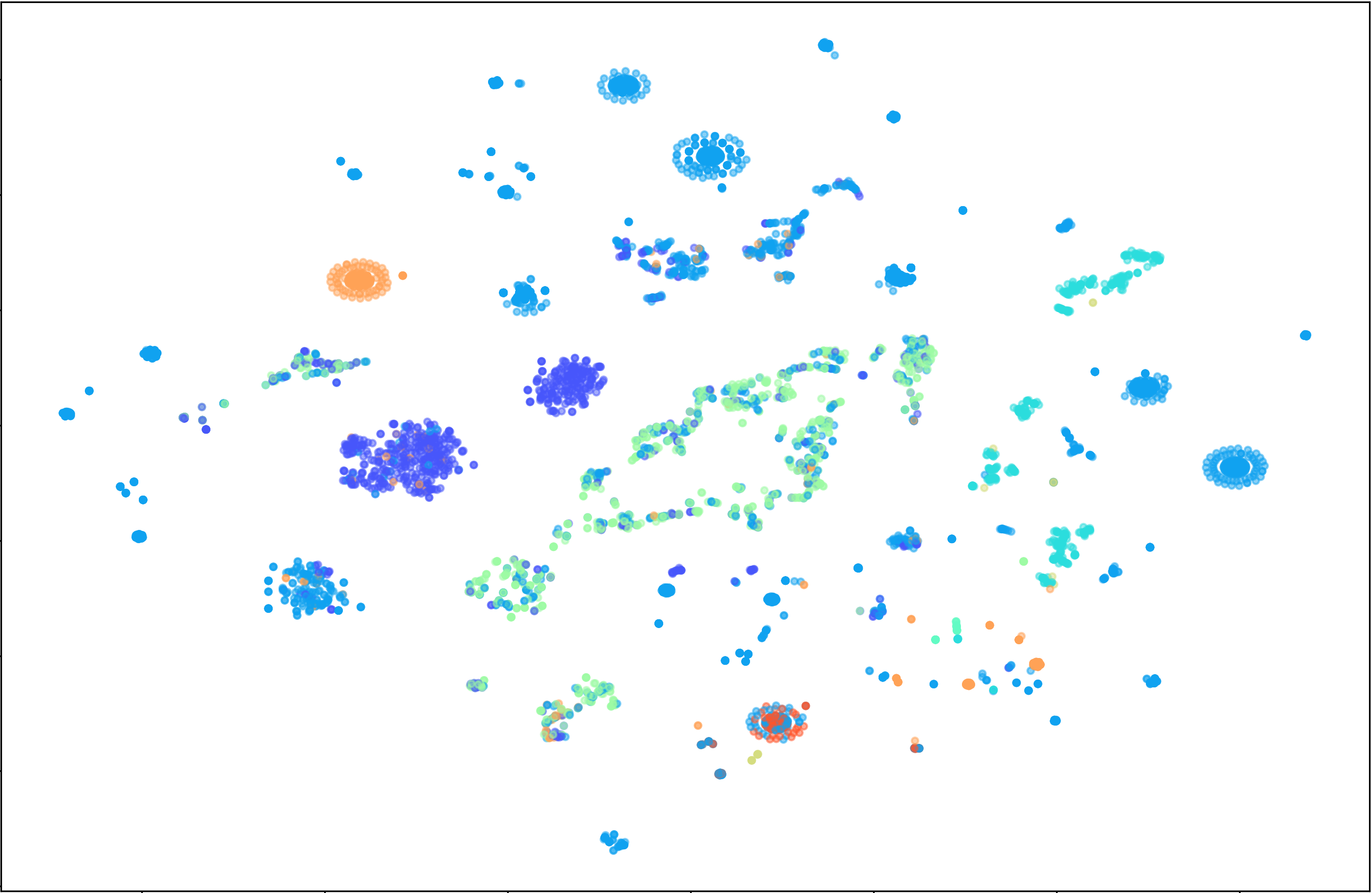}}
\caption{t-SNE projection of PELAT with K-means (purity=0.9054)}
\label{fig:kmeans}
\end{figure}

\begin{figure}[t]
\centering
\subfigure
[Colored by SOM cluster]
{\includegraphics[width=0.8\linewidth]{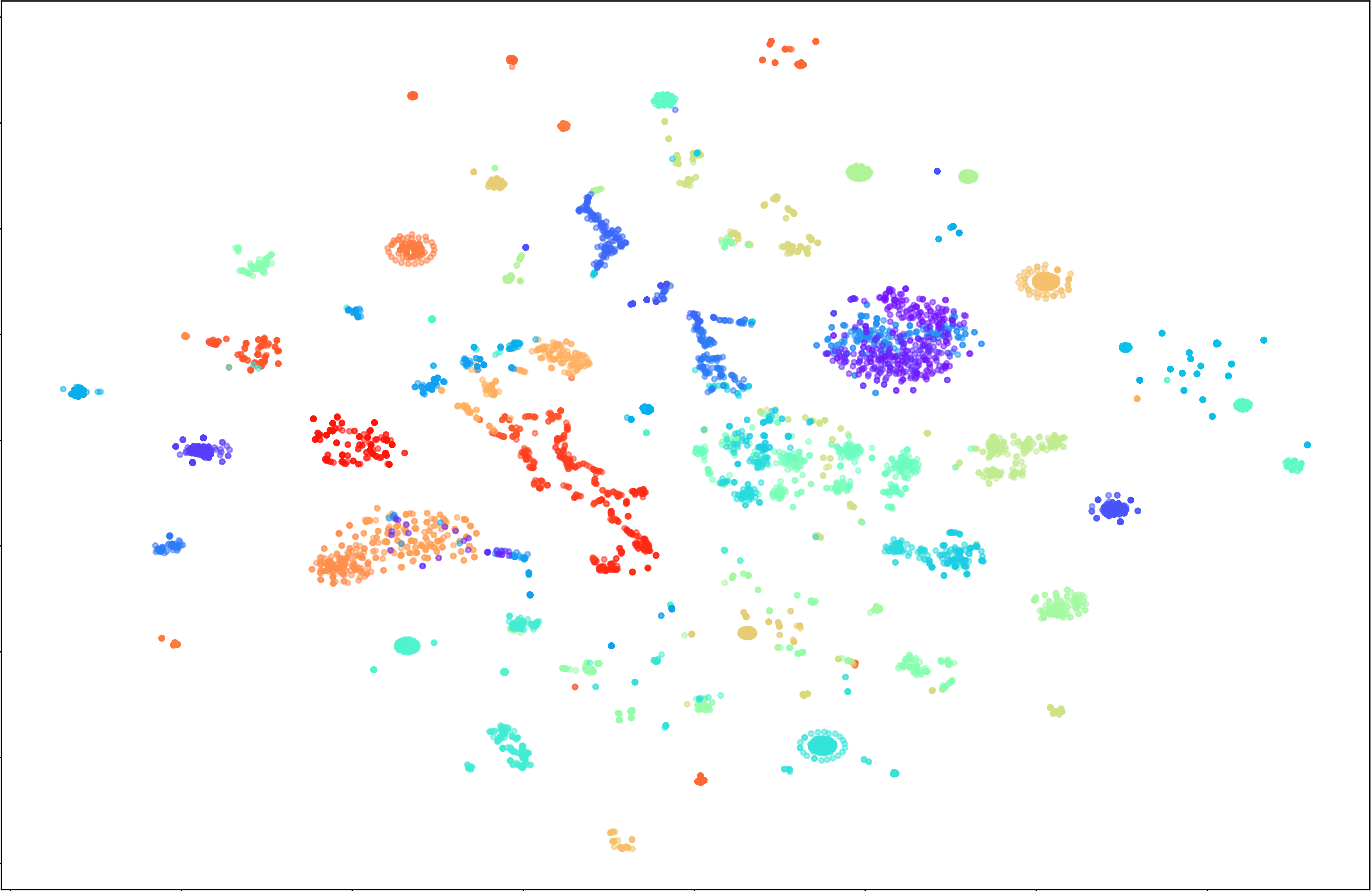}}
\subfigure[Colored by tactic label]{\includegraphics[width=0.8\linewidth]{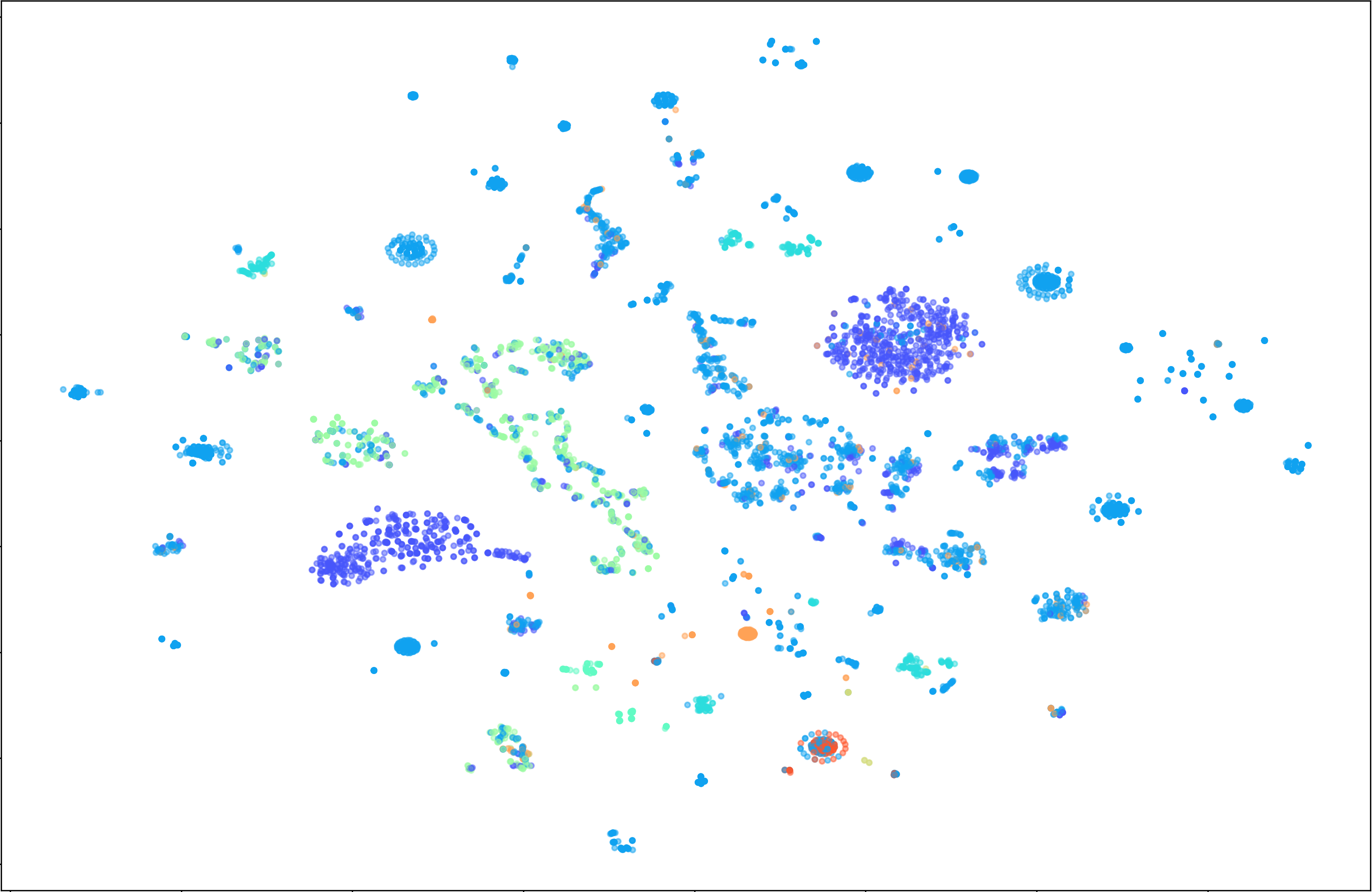}}
\caption{t-SNE projection of PELAT with SOM (purity=0.9045)}
\label{fig:som}
\end{figure}

To understand the effectiveness of our embedding, we compare the unsupervised clustering result of HTTP unique payloads in our embedding space with the supervised tactic labels on same payloads. We anticipate if the embedding is meaningful, same cluster of payloads should belong to same tactic. In our evaluation, purity are used to evaluate how close the clusters in the embedding space to the classes of labeled tactics.
We drop the duplicated payloads after the preprocessing. This remains 242,513 unique payloads. We make use of BERT$_{Base}$ uncased and set the maximum sequence length to 512 tokens with the embedding space of dimension 768. We reserve the \textit{CLS} vector as a payload representation. We implement both K-means clustering with K = 30 and self-organizing map (SOM) clustering with the initial of a $6\times6$ neuron map. To visualize the embedding result, we sample 200 dots from each cluster and plot t-SNE in a 2D space, which shows the variance of payload contents can be represented as different vectors on the embedding space. Purity's range is between 0 and 1. A cluster is pure if all data with the same classes are in the same cluster, the formula is used as follows:
\begin{equation}
Purity(\Omega, \mathbb{C})=\frac{1}{N}\sum\limits_k \max\limits_j | \omega_k \cap c_j |
\end{equation}
where $\Omega = \{\omega_1, \omega_2, ... , \omega_K \}$ is the set of clusters and $\mathbb{C} = \{c_1, c_2, ... , c_J \}$ is the set of tactic classes ($k \in K$, $j \in J$).
To verify consistency within the group, our K-means and SOM clustering result returns the purity of 0.9054 and 0.9045. We map tactics back to the embedding space (see Fig.~\ref{fig:kmeans} and Fig.~\ref{fig:som}).
Table~\ref{tab:purity} shows the purity of two different cluster algorithms with BERT and PELAT embedding. The higher purity indicates that the PELAT can not only represent a payload by a vector but also embed a packet payload with the knowledge of attack tactics. Furthermore, PELAT embedding is effective regardless of the clustering method used.

\begin{table}[tbp]
\small
\caption{The Purity of Clusters with Different Language Model Embedding}
\begin{center}
\begin{tabular}{lcc}
\hline
\textbf{Model}  & \textbf{K-means} & \textbf{SOM} \\
\hline
BERT$_{Base}$  & 0.7929 & 0.7903 \\
PELAT       & \textbf{0.9054} & \textbf{0.9045} \\
\hline
\end{tabular}
\label{tab:purity}
\end{center}
\end{table}

\subsection{Signature}
\begin{figure}[tbp]
\centerline{\includegraphics[trim = {0mm 0 0 0}, clip, width=0.9\linewidth]{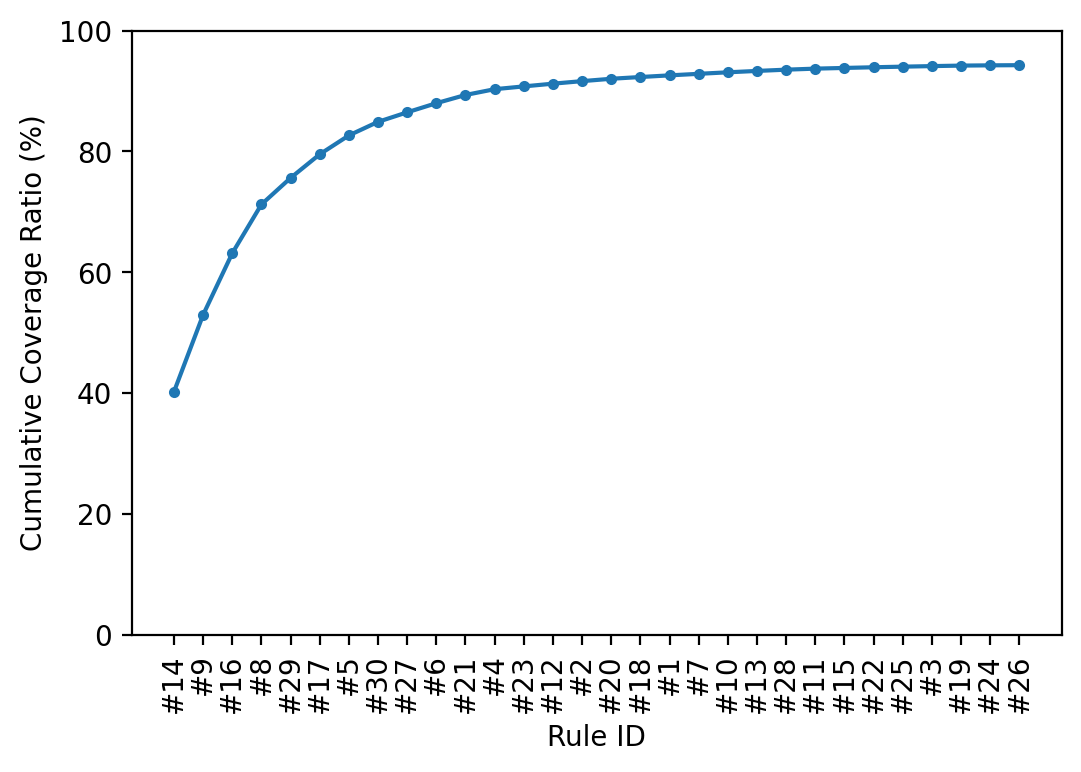}}
\caption{Rules coverage.}
\label{fig:coverage}
\end{figure}

Recently, Snort also adds a syntax to represent the TTP of the triggered packets. Only 3\% of the Snort default rules have predefined TTP. In our HTTP honeypot dataset, only less than 10 packets are triggered in a single day (around 300 thousand packets). To manually write Snort TTP rules for HTTP packets is impractical. We can leverage the proposed embedding method to automatically label a packet with its TTP. However, Snort rule needs content signatures rather than embedding vectors. We apply Term Frequency Inverse Document Frequency (TF-IDF) on every cluster to determine the relevance of signatures to the payload corpus and adopt tokens with high weight in TF-IDF to construct a Snort rule for a cluster. In this case, we can efficiently label the TTP of packets using Snort.
Furthermore, we discover a long tail phenomenon in the honeypot datasets, in which a large number of packets belong to only a few clusters. We notice that most malicious activities try to figure out login pages, post sexual or pharmaceutical ad comments, or form brute force attacks so that the attack behaviors are similar. Therefore, we take the advantage of the long tail phenomenon and we only need to construct a few Snort rules to cover high proportions of honeypot packets.
In our case, we construct 30 Snort rules for the 30 largest clusters by referencing their TF-IDF. They can cover 94\% of all the HTTP packets in our dataset (see Fig.~\ref{fig:coverage}). In addition, we can calculate the centroid of each cluster to represent the vector signature of this cluster. Both types of signatures can be used by firewalls or intrusion detection systems for security checking.

\subsection{Semi-supervised Learning}
\label{sec:semi}
By using Snort rules generated in the previous section, we trigger the total of 125,716 unique packet payloads with 6 tactics: Reconnaissance, Initial Access, Execution, Persistence, Credential Access, Discovery, also a Non-attack class for the packets that do not belong to attacks (see the ``Semi-sup. Learning'' column in Table~\ref{tab:dataset}). However, those packets are triggered by exact matching Snort content signatures which come from the TF-IDF tokens in the clusters. 
We then use them as the training dataset for the latter semi-supervised learning to train a multi-label classifier that accepts a packet vector and outputs its tactic. In this case, the classifier can deal with a small variant of packet payload and can still output its corresponding tactic by learning the relationship between the packet vector and its tactic. Hence, such an approach can be more flexible without the limitation of human-generated rules and the content-based exact matching approach.
We set a batch size of 16 and the maximum sequence length to 256 tokens according to the honeypot payload token distribution. We use the Adam optimizer with a learning rate of \texttt{2e-5}. The dataset is split into training, test, and validation sets with the ratio of 85\%, 10\%, and 5\%. After training the model, we predict the rest of the 29,776 unique packet payloads that Snort did not trigger. Since it is a multi-label classifier, if the probability of any label is less than 20\%, then we consider the tactic of this packet to be set to ``Unknown.'' If the ``Non-attack'' label is in the top $k$ (2) probability, then we consider there is no evidence this packet belongs to any of the attack tactics. If the probability difference between the highest and the second-highest tactic is within 10\%, the packet has two tactic labels.
In our semi-supervised learning process, all the unique packets are assigned with their tactics.
The predicted tactic result is shown in Table~\ref{tab:inferday} ``Semi. Test'' column.

\begin{table}[tbp]
\small
\caption{Tactic Samples Used in Different Learning Process}
\begin{center}
\begin{tabular}{lrrr}
\hline
   & \textbf{CTI} & \textbf{Semi-sup.} & \textbf{Transfer} \\
   & \textbf{Doc} & \textbf{Learning} & \textbf{Learning} \\
\hline
Reconnaissance (Rec) & 68 & 67,660 & 83,686 \\
ResourceDevelopment & 69 & - & - \\
InitialAccess (IA) & 72 & 12,717 & 16,026 \\
Execution (Exe) & 68 & 353 & 1,338 \\
Persistence (Per) & 247 & 89 & 136 \\
PrivilegeEscalation & 227 & - & - \\
DefenseEvasion & 456 & - & - \\
CredentialAccess (CA) & 180 & 17,313 & 19,845 \\
Discovery (Dis) & 90 & 90 & 1,836 \\
LateralMovement & 75 & - & - \\
Collection & 96 & - & - \\
C\&C & 91 & - & - \\
Exfiltration & 33 & - & - \\
Impact & 44 & - & - \\
Non-attack (Non) & - & 27,599 & 32,937 \\
Unknown (Unk) & - & - & 6 \\
\hline
Unique & 1,417 & 125,716 & 155,492 \\
\hline
\end{tabular}
\label{tab:dataset}
\end{center}
\end{table}

\subsection{Transfer Learning}
To perform the transfer learning, we adopt the PELAT model (trained in Section \ref{sec:languagemodelCTI}) and fine-tune it with all the data populated in Subsection \ref{sec:semi} (including the training data of semi-supervised learning 125,716 and the inferencing data of 29,776). In this case, we can transfer the knowledge learned from CTI reports into the knowledge of classifying packets to tactic categories.
We fine-tune PELAT model with three epochs and use the Adam optimizer learning rate \texttt{2e-5}, dropout probability of  0.1. According to the payload token distribution, the maximum sequence length is set to 256 tokens, same in Section \ref{sec:semi} with a batch size of 16.
The dataset is split into training, test, and validation sets with the ratio of 85\%, 10\%, and 5\% (see the ``Transfer Learning'' column in Table~\ref{tab:dataset}). We add a sigmoid non-linear classifier to predict eight classes which include six tactics (Reconnaissance, Initial Access, Execution, Persistence, Credential Access, Discovery) and Non-attack and Unknown classes. Table~\ref{tab:score} shows the test dataset 
scoring metrics. It shows that the model with transfer learning has good performance on classifying tactics. 


\begin{table}[tbp]
\small
\caption{Testing on PELAT Transfer Learning}
\begin{center}
\begin{tabular}{lrrrr}
\hline
\textbf{Class}  & \textbf{Prec.(\%)} & \textbf{Rec.(\%)} & \textbf{F1(\%)} & \textbf{Supp.} \\
\hline
CA & 100.0 & 99.94 & 99.97 & 1,817 \\
Dis & 100.0 & 100.0 & 100.0 & 184 \\
Exe & 94.74 & 81.20 & 87.45 & 133 \\
IA & 99.80 & 99.80 & 99.80 & 1,524 \\
Non & 99.78 & 99.97 & 99.87 & 3,133 \\
Per & 100.0 & 100.0 & 100.0 & 11 \\
Rec & 99.74 & 99.85 & 99.79 & 8,000 \\
Unk & 0.00 & 0.00 & 0.00 & 1 \\
\hline \hline
Micro Avg & 99.75 & 99.71 & 99.87 & 14,803 \\
\hline
\end{tabular}
\label{tab:score}
\end{center}
\end{table}

\subsection{Inference on Packets with Transfer Learning}
We make inference on two the other days from our collected honeypot dataset that the total inbound HTTP packets are 227,109 and 227,006. The tactic inferencing results by PELAT are shown in Table~\ref{tab:inferday} column ``Day A'' and ``Day B''. It shows that the transfer learning which only learns 1,417 MITRE ATT\&CK documents can successfully infer 86.96\% and 88.71\% of attack tactics (exclude non-attack packets identified by PELAT) used by 227,109 and 227,006 packets, respectively. It is more efficient to learn the attack knowledge from the well-labeled CTI documents for inferencing attack tactics. It reduces the burden of massive and manual labeling on several hundreds of thousands of packets.

\begin{table}[tbp]
\small
\caption{Inference Result on Different Learning Process}
\begin{center}
\begin{tabular}{lrrr}
\hline
\textbf{Class} & \textbf{Semi. Test} & \textbf{Day A} & \textbf{Day B} \\ \hline
Rec            & 15,883        & 153,124   & 161,858 \\
Non            & 5,338         & 29,609    & 25,628  \\
IA             & 3,239         & 8,415     & 1,908   \\
CA             & 2,532         & 26,101    & 28,061  \\
Dis            & 1,739         & 740       & 1,054   \\
Exe            & 779           & 4,491     & 2,982   \\
Rec, Exe       & 136           & -         & -       \\
IA, Exe        & 66            & 3,898     & 4,594   \\
Per            & 47            & 739       & 921     \\
Unk            & 6             & 0         & 0       \\
Exe, Dis       & 4             & -         & -       \\
Rec, IA        & 4             & -         & -       \\
Rec, Dis       & 3             & 1         & -       \\ \hline
Total          & 29,776        & 227,109   & 227,006 \\
\hline
\end{tabular}
\label{tab:inferday}
\end{center}
\end{table}

\section{Discussion}
\label{sec:discussion}
\subsection{Findings and Thoughts}
This study is the first research that adopts transformer-based language model to embed text-based network packets to vector-based representation for the latter analysis. Especially, we adopt transfer learning technique to learn the knowledge from CTI reports and transfer it to predict the attack tactic for network packets.

Some findings from our experiments are listed as follows.

\begin{enumerate}[a)]
\item The proposed fine-tuned PELAT model (with CTI knowledge ) outperforms the vanilla BERT model when performing CTI tactic classification. Such approach is named \textit{BERT Experts} \citep{bertexpert}. The outperformed results show that we indeed `add' the CTI domain to the BERT and could align more closely with the target task. 


\item The reason why HTTP packets in our experiments can be clustered is that the packet data is obtained from certain hidden honeypots built in the ISPs. Only the attack traffic that perform random IP scanning could possibly hit the honeypots. Usually, the these attacks are automated programs, such as worm and bots. It causes the data distribution follow a power law \citep{powerlaw}, i.e., most of the attack traffic belong to a small portion of the cluster. The automated attacks result in such long-tail data. As shown in Fig.~\ref{fig:coverage}, we can filter out these packets with a relatively smaller number of rules. It makes the analysis more efficient. There still exists certain small clusters in Fig.~\ref{fig:kmeans} and Fig.~\ref{fig:som}. We anticipate they are rare-seen attacks in the honeypot, such as APT \citep{apt}, and it could be a future research work to analyzing them.

\item The latent space of packet embedding demonstrates that the attack packets from honeypots are well-represented by our model and well-clustered under different clustering algorithms (as shown by the high purity of clusters). Table~\ref{tab:purity} shows that we can increase the embedding performance by the PELAT model. PELAT model is trained on CTI reports (i.e., text) from a similar domain (i.e., security article) to a similar task (attack tactic classification). It reveals that our approach of fine-tuning on the downstream task gives impressive results on security task \citep{bertexpert}.  We anticipate the embeddings can be used for further (security-related) tasks.

\item We can output the byte-based signature from TF-IDF and the vector-based signature from cluster centroid for each cluster in the latent space. Since the byte-based signature relies on string matching technique, the vector-based signature relies on distance measuring technique to determine whether matched or not. We anticipate that the vector-based signature is more flexible and general to identify certain attacks with a small variance in the packet payload. That is the distance increases by the small variance is still small enough without leaving a centroid too far away.

\item Table~\ref{tab:dataset} shows that we use `Semi-sup. Learning' data (unique packet payloads in 2 days) to train our model for tactic classification; while Table~\ref{tab:inferday} shows PELAT makes inference result to the packets collected from two other days. The classification result of two other days both have zero sample of `unknown (Unk)' class. It indicates that PELAT can at least predict one of the classes with high probability (in practice, larger than 0.5). It also reveals that the traffic in the honeypot is relatively homogeneous.

\end{enumerate}

\subsection{Limitations}
\begin{enumerate}[a)]
\item We use real-world attack data collected from honeypots in ISPs. It relies on the the interaction capability of the installed honeypots to collect high-quality attack traces. However, Amun, Cowrie, Dionaea, Glastopf are all low-interactive \citep{honeypot-low}, and they do not cover all the attack tactics defined by the MITRE. In this case, some of the tactics are not observed, although PELAT leans all tactics by CTI reports, the testing data does not include certain tactics (as the `-' shown in Table~\ref{tab:dataset}). However, if attack samples under all tactic phases are available and well-labeled, PELAT can fill in value of `-'.

\item Clearly, the number of CTI Doc in MITRE's website is unbalanced and some reports are messy (for example, an OCR-unfriendly PDF file or a HTML-tagged web page). The quality of the CTI report relies on the preprocessing methods applied to these messy files (including 1400+ reports with around 1.5 million words). We have tried our best to design our preprocessing phase. However, we anticipate that if high-quality of corpus is available, the accuracy of Table~\ref{tab:conparison} could be higher.


\item MITRE ATT\&CK TTP update continually. As a result, some techniques (or tactics) are removed, integrated, or renamed. Thus, we have to retrain our model to be consistent with the latest TTP definition.
\end{enumerate}

\subsection{Future Works}
\begin{enumerate}[a)]
\item Pre-train the model with cloze task \citep{bert} and permutation/rotation task \citep{xlnet} to learn the dialect of HTTP and apply more downstream tasks to improve PELAT for robustness and generalization.

\item Test our model on other ISP's datasets and long-span datasets for evaluating the robustness of the design.

\item Apply PELAT on other text-based protocols such as SMTP, and further, on binary protocols.



\item Propose a new embedding method based on PELAT by using a sequence of a packet (i.e., session). In this case, we can better understand the attack (or attack's intention) from the perspective of consecutive communication, rather than a single packet.

\end{enumerate}

\section{Conclusions}
\label{sec:conclusion}
This paper proposes a model, PELAT, to leverage transfer learning and a transformer-based language model to analyze text-based attack (HTTP) packets for attack tactic identification. The transfer learning mechanism can increase the knowledge of the language model by learning security reports to further create an embedding latent space that can vectorize a packet with respect to its corresponding attack tactic. The boosted accuracy and ROC of the PELAT for CTI tactic classification are 3.1\% and 17.1\%, respectively, compared to the BERT model. The embedding result in PELAT can achieve 0.9054 and 0.9045 of purity considering the true tactic label and clusters of our embedding space. The transfer learning result also shows that PELAT can predict the tactic for each and every packet in two new datasets (i.e., zero unknown class). We are the first study that combines the transfer learning and language model for network packet embedding. The model embeds the knowledge of English (from BERT) and attack lifecycle (from MITRE ATT\&CK CTI report). We anticipate this work can help the security expert dealing with the unstructured packets, and can leverage the embedding latent space of PELAT for further downstream security analysis.

\bibliographystyle{IEEEtran} 
\bibliography{cas-refs}






\end{document}